# Magnetic phase diagram of Ge$_{1-x-y}$(Sn$_x$Mn$_y$)Te multiferroic semiconductors: coexistence of ferromagnetic and cluster glass ordering

A. Khaliq,[1,a] S. Lewińska,[1] R. Minikaev,[1] M. Arciszewska,[1] A. Avdonin,[1] B. Brodowska,[1] V.E. Slynko,[2] A. Ślawska-Waniewska,[1] and L. Kilanski,[1]

[1]*Institute of Physics, Polish Academy of Sciences, Aleja Lotnikow 32/46, PL-02668 Warsaw, Poland*
[2]*Institute of Materials Science Problems, Ukrainian Academy of Sciences, Chernovtsy, Ukraine*

We report the structural and magnetic results of polar α–GeTe doped with Sn and Mn from $x = 0.185$ to $0.841$ and $y = 0.02$ to $0.086$, respectively. The magnetic results of Ge$_{1-x-y}$(Sn$_x$Mn$_y$)Te (GSMT) crystals identify Mn-clustering effect with scaling parameter, $R = 0.033$ for $x \approx 0.2$ and $y = 0.06$. The excessive Sn ions are assumed to drive the inception of short range ferromagnetic clusters. For the crystals revealing glassy magnetic behavior, the irreversibility temperature, $T_{irr}$ shifts at high dc magnetic field which is well described by de Almeida-Thouless equation, $\delta T_{irr} \propto H^{\Phi/2}$ yielding Φ values of 1.55 and 1.7 that validates the formation of Mn clusters. The spin relaxation time, $\tau_0 \sim 10^{-9}$ s, activation energy, $E_a/k_B \sim 5\, T_F$ where $T_F$ is freeing temperature and Vogel-Fulcher temperature, $T_0 \sim T_F$ also signify intracluster interactions. Also, the Mn-hole magnetic exchange constant, $J_{pd}$ drops from 0.24 eV for the samples with smaller Sn content of $x \approx 0.2$ to 0.16 eV for Sn rich alloy with $x \approx 0.8$. Consequently, we present a magnetic phase diagram for GSMT bulk crystals as a function of Mn($y$) ions.

## I. INTRODUCTION

The development of new spintronic materials based on randomly distributed Mn ions in a host lattice has seen numerous efforts both experimentally and theoretically [1]. In contrast to conventional electronic materials centered on electronic charge only, spintronics offer an integration of tunable semiconducting nature and spin degree of freedom in a single system [1–3]. In order to develop functional spintronic materials, diluted magnetic semiconductors (DMSs) belonging to group II–VI [4,5], III–V [6,7] and IV–VI [8,9] have been widely studied doped with magnetic elements. Until now, DMSs have seen pronounced development in elevating its critical temperature, $T_C$, such as (Ga,Mn)As epilayers with about 11% − 13% of Mn achieved $T_C = 185$ K by Olejník *et al*., [10]. Chen et al. further enhanced the $T_C$ value to about 200 K in (Ga,Mn)As thin film using nanostructure engineering [11]. However, previously obtained results show that functional ferromagnets are still far from their anticipated applications due to the low $T_C$ values. For this reason, DMSs have been a dynamic research topic in condensed matter physics in order to advance their functional capabilities.

Among several compound semiconductors (SCs), IV–VI DMSs offer a wide range of possibilities for instance higher solubility of transition metals [12], possible tunability of ferroelectric features and Rashba effect in spin–bands [13], entanglement between magnetic order and spin orbit interaction (SOI) [14] and spin-light conversion due to higher-order Dresselhaus SOIs in systems with broken symmetry such as α–GeTe [15]. Due to the presence of broken symmetry and narrow band gap, α–GeTe and its derivatives provide a reliable playground to control spin-texture via ferroelectric polarization and exploit giant Rashba spin splitting [13,16]. From the ferromagnetic (FM) perspective, Mn doped α–GeTe has demonstrated considerable enhancement of $T_C$ up to 200 K [17] since its emergence in 1966 by Rodot *et al*. [18]. The $T_C$, coercivity and magnetic features in DMSs could be manipulated by varying the content of the doped magnetic element [19]. Linear correlation between Mn concentration and $T_C$ was shown by Park *et al*., for Mn$_{1-x}$Ge$_x$ where $T_C$ was raised from 25 K to 116 K as a function of Mn content [20].

The nature of magnetic interactions in Mn based alloys has been studied in the absence of holes in the sample for different compound SCs like In$_{1-x}$Mn$_x$As [21] and fully carrier compensating (Ga,Mn)As [22] displaying antiferromagnetic coupling between Mn ions. Apart from this, the FM order induced in Mn doped IV–VI crystals for example Sn$_{1-x}$Mn$_x$Te and Ge$_{1-x}$Mn$_x$Te which have charge concentration of $p \approx 10^{21}$ cm$^{-3}$, is believed to originate from Ruderman−Kittel−Kasuya−Yosida (RKKY) indirect exchange interaction mediated by hole carriers [23,24]. Previous studies indicate that DMSs crystals exhibit a wide spectrum of magnetic features depending on the doping content and consequently the presence of free charge carriers. Crystals for example Pb$_{1-x}$Mn$_x$Te possessing low charge carrier concentration manifested a paramagnetic (PM) state [25], but also a spin−glass (SG) ordering was observed for Mn ≤ 0.2 [26] below 1 K, and also FM phase for very high charge carriers [27]. These studies suggest a broad range of magnetic ordering and interactions could be investigated in DMSs both as a function of doping and temperature.

In this paper, we present GSMT IV–VI quaternary crystals with diamagnetic (DM) and PM doping $x = 0.18–0.84$ and $y = 0.02–0.09$, respectively, a continuation of previous work [28–30]. The contents of Sn/Mn ions are considerably varied to investigate its influence on the structural and magnetic properties of α–GeTe based DMSs. Precisely, the magnetic interactions in high Sn region are of particular interest in this work. Since alloys with broad Mn content have been previously studied, the variation in Sn between $x$ ~0.4 and ~0.8 offer possibilities to tune the magnetic interactions in high quality GSMT multiferroics (MF). We show that such a heavy doping of Sn ions have substantial impact on the magnetic ordering in the present alloys. Specifically, the impact of high Sn doping to alter the nature of frustrated magnetic state is discussed. Prior to these results, earlier works on GSMT samples showed different magnetic orderings e.g. canonical SG [28,30] and FM like features for some crystals [29]. The present project extends far beyond Ref. 28-30 by focusing on the study of the entire range of chemical compositions of Sn from 0 to 1. Therefore, this work is an attempt to obtain an inclusive report of the magnetic interactions present in GeMnTe-SnMnTe mixed crystals as a function of both DM Sn and PM Mn ions. Consequently, this work is presented to be perceived as an extensive analysis of magnetic properties of GSMT as our earlier work was limited only to a narrow range of chemical compositions. Here, the frustrated magnetic state reveals a distinct and collective behavior rather than individual spins. This manifestation of spin dynamics is identified to originate from clustering of Mn ions. In comparison with Ref. 28-30, the intracluster interactions are particularly noticed for samples with high Sn contents i.e. ~0.4 – ~0.6 however; similar Mn concentration. This probably indicates that the different lattice dimensions of Sn rich crystals of $Sn_{1-y}Mn_yTe$ compared to $Ge_{1-y}Mn_yTe$ are responsible for Mn clustering effect. We assume that the diverse magnetic behavior of GSMT and formation of clusters in particular is attributed to altering of inter-Mn distances. It is also seen that the $\chi_{AC}(T)$ at different $f$ values depict distinct magnetic behavior from the SG state noticed earlier [28,30] which is driven by clusters in the present samples.

In this sense, the magnetic susceptibility as a function of temperature, $\chi_{AC}(T)$ results were obtained for magnetically glassy samples at several different frequencies, $f$ of the alternating applied magnetic field. A typical for SG state, the shift of $\chi_{AC}(T)$ maximum with the $f$ values was registered. Furthermore, the analysis of the temperature dependence of magnetization, $M(T)$, showed that the anomaly in zero field-cooled (ZFC) curves moved towards lower temperature as the magnitude of applied dc magnetic field increased. Our results show that such a behavior of disordered magnetic state agrees with de Almeida–Thouless (AT) line following a power law $\delta T_{Irr} \propto H^{\Phi/2}$. Moreover, the variation of $T_F$ was analyzed with critical slowing down and Vogel-Fulcher law which identified an FM–like ordering and disordered state in coexistence. Finally, we present a magnetic phase diagram for GSMT crystals as a function of Mn content which alters the nature of magnetic ordering in the studied alloys.

## II. SAMPLE PREPARATION

The bulk crystals of GSMT MF were grown by modified vertical Bridgman method as described in our previous work [28–30], originally these modifications were designed by Aust and Chalmers in order to obtain aluminum crystals of improved structural quality [31]. The solubility of Mn ions in $Ge_{1-x}Sn_xTe$ lattice and homogeneity of the final samples was carefully achieved using improved growth conditions. The growth process of the desired GSMT quaternary crystals was initiated with GeTe, SnTe, and $MnTe_2$ as precursors which were previously prepared by high-temperature synthesis from high-purity elemental components. Prior to the crystals' growth, the modified conditions developed by Aust and Chalmers were introduced. In comparison with metallic manganese, melting point of 1244 °C [32], we used $MnTe_2$ which enhanced the dissolution of Mn into the melt due to its relatively low melting point of 941 ± 3 °C [33]. For the directional crystallization of the ingot at high temperature, the inner surface of the quartz ampoules was covered with thin semitransparent graphite layer which minimized interaction of the melt with quartz as well as adhesion of the synthesized ingot to the walls of the ampoules. During the directional crystallization, the ampoules were drawn at a speed of 1.5 mm/hr through the furnace whereas temperature gradient was kept at 30–40 °C at the crystallization front. Prior to execution of the growth process, an additional heat source was placed in the crystallization zone to achieve a radial temperature gradient in addition to the established longitudinal gradient. With these changes, we were able to move the angle of the interface of solid and liquid states by ≈ 15° relative to the horizon line. This deviation improved melt mixing and reduced the crystal blocks to one in as-grown cylindrical ingot. The method described above incorporated Sn and Mn ions into GeTe lattice along the direction of growth axis. During crystal growth, the x-ray fluorescence technique showed that the Sn and Mn concentrations gradually increased and decreased, respectively.

## III. EXPERIMENTAL TECHNIQUES

Energy dispersive x–ray fluorescence (EDXRF) technique using Tracor x–ray Spectrace 5000 EDXRF spectrometer was used to find out the chemical compositions of GSMT alloys. The room temperature crystal structure measurements were made using high resolution multipurpose XPert Pro MPD x–ray diffractometer (HRXRD) of conventional tube x–ray source equipped with $CuK_{\alpha 1}$ radiation of wavelength, $\lambda$ = 1.5406 Å. The temperature dependence of the magnetic susceptibility, $\chi_{AC}(T)$, of the bulk crystals of GSMT was analyzed using Lakeshore 7229 AC susceptometer within the temperature range $T$ = 4.5 K to 120 K, and at frequencies, $40 \leq f$ (Hz) $\leq 10000$. Magnetization as a function of magnetic field, $M(H)$ curves, were measured using Lakeshore 7229 magnetometer equipped with superconducting magnet which enables measurements at magnetic fields up to $B$ = 9 T, whose principle of operation is based on the Weiss extraction method. Furthermore, the ZFC and FC magnetization measurements as a function of temperature, $M_{ZFC-FC}(T)$ were performed using a vibrating sample magnetometer (VSM) as an option of a commercial Physical Property Measurement System platform (PPMS, Quantum Design). The $M_{ZFC-FC}(T)$ measurements were executed at dc magnetic field in the range, $10 \leq H_{dc}$ (Oe) $\leq 200$.

## IV. STRUCTURAL CHARACTERIZATION

The chemical concentrations of the incorporated Sn and Mn ions into α–GeTe lattice were determined by means of energy dispersive x–ray fluorescence (EDXRF) method using Tracor x–ray Spectrace 5000 EDXRF spectrometer. The deduced chemical compositions were obtained with less than 10% uncertainty as compared to the nominal values of Sn and Mn ions in the entire series of doped GSMT crystals. The uncertainties between the calculated and nominal chemical values are similar to GSMT alloys presented earlier with different Sn and Mn contents [28,29]. For the structural and magnetometric measurements, each disc-shaped crystal was further cut into 1×1×8 mm$^3$ bars. The chemical compositions and crystallographic parameters are presented in Table I deduced from the EDXRF and XRD measurements. The incorporation of Sn and Mn ions into α–GeTe lattice is projected to influence its crystal structure which holds a rhombohedral symmetry (JCPDS, no.47−1079) [34] below $T \approx 720$ K [15]. For this reason, the room temperature (RT) HRXRD measurements in the range $2\theta = 20°–150°$ were taken in order to analyze the influence of doping on the crystal structure of α–GeTe.

In previous study in which the impact of small doping content of Sn and Mn was investigated, the crystals showed a rhombohedral symmetry (R3m) for low contents [30] of Sn and Mn however, a mixed phase of cubic rock salt structure, Fm-3m (JCPDS, no. 54–0498) [35] and R3m for intermediate doping amounts [30]. Owing to the higher doping solubility in IV–VI materials, the amounts of Sn and Mn impurities were further increased into α–GeTe lattice. In the present work, for the crystals with lower doping contents of $x \sim 0.4$ and $y = 0.02$, the two phase structure persists at RT, presented in Fig. S1(a) of the supplementary figures. As the Mn content is increased to $y = 0.052$ for the same Sn content, the coexistent R3m+Fm-3m phase changes into pure Fm-3m symmetry for the entire series of the heavily doped crystals, Fig. S1(b-h).

**TABLE I:** The chemical compositions deduced from EDXRF analysis and structural parameters are summarized in the following table, $x$ and $y$ show the molar concentrations of Sn and Mn ions, $a$ and $c$ represent the lattice parameters along $a$ and $z$ axis for rhombohedral symmetry, the crystal structure and space groups obtained from the HRXRD measurements are provided.

| $x \pm \Delta x$ | $y \pm \Delta y$ | $a$ [Å] | $c$ [Å] | crystal structure | space group |
|---|---|---|---|---|---|
| 0.185 ± 0.01 | 0.060 ± 0.006 | 8.460 ± 0.002 | 10.59 ± 0.02 | rhombohedral/ | R3m |
| | | 5.994 ± 0.002 | – | cubic | Fm-3m |
| 0.386 ± 0.03 | 0.020 ± 0.002 | 8.581 ± 0.002 | 10.67 ± 0.02 | rhombohedral/ | R3m |
| | | 6.161 ± 0.002 | – | cubic | Fm-3m |
| 0.405 ± 0.04 | 0.052 ± 0.005 | 6.097 ± 0.002 | – | cubic | Fm-3m |
| 0.406 ± 0.04 | 0.072 ± 0.007 | 6.097 ± 0.002 | – | cubic | Fm-3m |
| 0.545 ± 0.05 | 0.030 ± 0.003 | 6.098 ± 0.002 | – | cubic | Fm-3m |
| 0.593 ± 0.06 | 0.062 ± 0.006 | 6.182 ± 0.002 | – | cubic | Fm-3m |
| 0.641 ± 0.06 | 0.086 ± 0.009 | 6.192 ± 0.002 | – | cubic | Fm-3m |
| 0.724 ± 0.07 | 0.040 ± 0.004 | 6.198 ± 0.002 | – | cubic | Fm-3m |
| 0.797 ± 0.08 | 0.072 ± 0.007 | 6.254 ± 0.002 | – | cubic | Fm-3m |

Kriener et al., who worked on bulk $Ge_{1-x}Mn_xTe$ polar SC, showed that the R3m symmetry of α–GeTe switches to Fm-3m structure at about 12% of Mn doping [36]. The RT ferroelectric to paraelectric phase transition in this work occurs at about $x \approx 0.4$, $y=0.052$ which is compositionally different from the compounds studied in [36] due to the presence of Sn in the quaternary alloy. In the compositions with higher Sn content, the R3m α–GeTe becomes less dominant in the presence of SnTe crystal which has an Fm-3m phase at RT, therefore a ferroelectric to paraelectric phase transition is observed.

## V. EXPERIMENT AND RESULTS

Fig. 1(a,b,c) presents results of the AC magnetic susceptibility, $\chi_{AC}$ obtained at an alternating magnetic field of amplitude $H_{AC} = 10$ Oe and frequency, $f = 625$ Hz in which both the real, $Re(\chi_{AC})$, and imaginary, $Im(\chi_{AC})$, parts of $\chi_{AC}$ were measured. For accurate results, the holder contribution was subtracted from the $Re(\chi_{AC})$ and $Im(\chi_{AC})$ dependences of the samples. For the sample with $x \sim 0.4$, both the $Re(\chi_{AC})$ and $Im(\chi_{AC})$ parts manifest cusp like shapes for $y = 0.052, 0.072$ as presented in Fig. 1(a). For further analysis of the $\chi_{AC}$ results, the maximum value for each $Re(\chi_{AC}(T))$ cusp was calculated by fitting the curves with the Gauss function which yielded $T = 7.1$ K and 10.5 K for $y = 0.052$ and 0.072, respectively. As obvious from the figure, the maxima of both $Re(\chi_{AC})$ and $Im(\chi_{AC})$ parts occur at the same temperature. However, the shape of $\chi_{AC}$ cusp for each crystal is different, a broad peak for $y = 0.072$, whereas a sharp one for $y = 0.052$. The magnitude and broadness of the cusps increase with higher $y$ contents. Anomalies, like maxima, in the $\chi_{AC}(T)$ curves are indicative of different types of magnetic ordering which could occur as a function of impurities in the crystals. A maximum in the $\chi_{AC}(T)$ graph of semimagnetic SCs is usually associated with a glassy magnetic ordering or superparamagnetic state, but it might also represent a transition between a PM and FM phase. [37-39] However, for the samples with $x = 0.4$ and $y = 0.072$ and 0.052 the rapid decrease of the $\chi_{AC}(T)$ dependence below the peak temperature rather suggests the former state. (The deeper analysis of the origin of the maxima will be presented further in the text.) In contrast, the sample with $x \sim 0.4$ and $y = 0.02$ displays a very small non-zero $\chi_{AC}(T)$ value only near $T = 4.5$ K which does not manifest any obvious magnetic ordering in this temperature range. Such a result seems to be correct as it is the sample with the lowest content of the Mn ions. In the situation when the concentration of Mn ions is very low, and these ions are well separated in the whole volume of the sample, a system may behave as PM.

In Fig. 1(b), $\chi_{AC}(T)$ results are shown for the crystals with higher Sn content of $x \sim 0.6$ which were measured under the same conditions as the samples in Fig. 1(a). Here, the samples with highest Mn content, $y = 0.062$ and 0.086 illustrate a plateau-shaped susceptibility at low temperatures with a sharp decrease to 0 near the temperatures named $T_C$. The $T_C$ temperature, for those GSMT crystals which required, were estimated by taking the double derivative of the $Re(\chi_{AC})$ with temperature, $\partial^2 Re(\chi_{AC})/\partial T^2 = 0$. For the samples with $x \sim 0.6$ and $y = 0.062$ and 0.086, the $T_C$ values are equal to 7 K and 12 K, respectively. The $Im(\chi_{AC})$ part is considerably smaller for the samples exhibiting plateau-shaped $\chi_{AC}(T)$ as compared to those showing cusps. For example, for $y = 0.086$, the maximum value of $Im(\chi_{AC})$ part is about 11% of the $Re(\chi_{AC})$ which is ~29% of $Re(\chi_{AC})$ for $x \sim 0.4$ and $y = 0.052$. On further increasing Sn content to $x \sim 0.8$, the Re($\chi_{AC}$) relations exhibit similar shapes to the ones seen for $x \sim 0.6$, however with negligible contribution of the Im($\chi_{AC}$) part.

We further analyze the magnetic $\chi_{AC}(T)$ by assessing the consistency of inverse dependences of Re($\chi_{AC}(T)$) with the modified Curie–Weiss (CW) law as expressed by eq. 1.

$$\text{Re}(\chi_{AC}) = \frac{C}{T-\theta} + \chi_{dia} \quad (1)$$

The modified CW law fitting was attempted in the temperature range between $T = 40 - 120$ K for the PM part of the results. It is worth mentioning that the use of CW law is implemented in the high temperature region above the transition temperature, $T_C/T_g$ of all samples. This means that all the samples are in the same spin-disordered or PM state from $T = 40 - 120$ K. Therefore, this should avoid the misperception that the CW law is presented only for samples showing PM state. Here, $C$ represents CW constant, $\Theta$ is paramagnetic CW temperature and $\chi_{dia}$ denotes the diamagnetic part of the host lattice GeTe, which contributes $\chi_{dia} = -3 \times 10^{-7}$ emu/g to the CW fits. The solid lines in Fig. 1(d,e,f) show a good fit of the modified CW law to the experimental data. The values of $C$ obtained by fitting Re($\chi_{AC}(T)$)$^{-1}$ with eq. 1 were used to calculate the effective magnetic moment by using, $\mu_{eff} = \sqrt{\frac{3k_B C}{N_A}}$, here $k_B$ denotes Boltzmann constant, whereas $N_A$ is Avogadro's number. The $\mu_{eff}$ values were calculated in the units of Bohr magneton, $\mu_B$, see in Table II. The tendency of $\mu_{eff}$ as a function of Mn content shows a linear correlation for the entire series of the samples with the exception of $x \sim 0.6$, $y = 0.062$ and $x \sim 0.8$, $0.072$ showing a slight deviation. This deviation probably occurs due to different Sn content which leads to dissimilar lattice constants and therefore unequal inter-Mn distances. Variation of $\mu_{eff}$ as a function of Mn content is shown in Fig. S2.

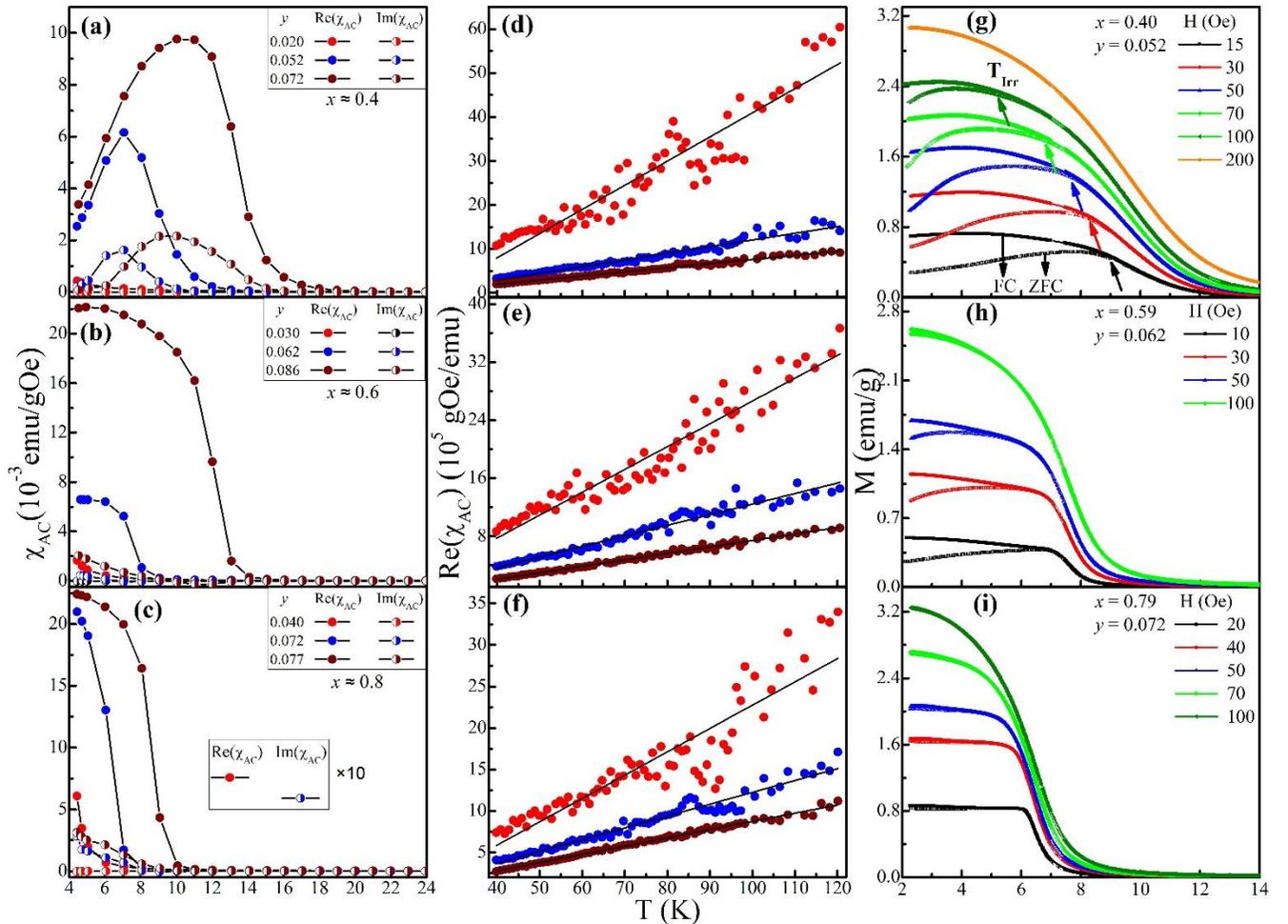

**FIG. 1.** (a,b,c) The Re($\chi_{AC}$) and Im($\chi_{AC}$) components of the temperature dependent AC magnetic susceptibility as a function of $x$ and $y$, measured at an alternating magnetic field, $H_{AC} = 10$ Oe and frequency, $f = 625$ Hz. In (c), the Re($\chi_{AC}$) and Im($\chi_{AC}$) parts of $y = 0.04$ and $y = 0.072$, respectively, were magnified by an order of 10 to express a readable graph. (d,e,f) The corresponding results of inverse of Re($\chi_{AC}$) fitted with the modified Curie-Weiss law in the PM region. Solid lines in (d,e,f) represent modified CW law fitting and (g,h,i) illustrate the temperature dependent ZFC-FC magnetization of selected samples from (a,b,c) measured for several dc magnetic field magnitudes.

Here, the spin only ground state of Mn$^{+2}$–3d$^5$ shell makes a total spin moment, $S = 5/2$. Besides the analysis of the PM part, the CW law fitting is an appropriate way to estimate the values of $\Theta$ from the experimental data which consequently reveals the nature of magnetic interactions between the magnetic ions in the lattice. For this reason, C and $\Theta$ were used as fitting parameters in eq. 1. The constant C in eq. 1 is expressed as,

$$C = \frac{N_0 g^2 \mu_B^2 J(J+1) x_\theta}{3k_B} \quad (2)$$

Here, $N_0$ is the number of cation sites per gram [40], $g$ is the [41], $\mu_B$ is Bohr magneton and $x_\theta$ is an effective Mn content. Since the $\chi_{AC}(T)$ results revealed different shapes for the studied crystals, therefore, $M(T)$ measurements at different dc magnetic field magnitudes were performed to further assess the nature of the magnetic interactions. Lande spin splitting g–factor which is equal to 2 for $Mn^{+2}$ ions

**TABLE II:** The Sn content $x$, Mn content $y$, critical temperature $T_C$, Curie Weiss−temperature $\Theta$, Curie−Weiss constant $C$, spin or cluster-glass temperature $T_F$, remanence magnetization $M_R$, coercive field $H_C$, scaling parameter $R$, saturation magnetization $M_S$ at $T = 4.5$ K, the dynamical exponent in Almeida−Thouless equation $\alpha$ and $\mu_{eff}$ is the effective magnetic moment in the units of Bohr magneton, $\mu_B$.

| $x$ | $y$ | $T_C$ [K] | $\Theta$ [K] | $C$ ($10^{-4}$) [emu.K/g] | $T_F$ [K] | $M_R$[emu/g] $T = 4.5$ K | $H_C$ [Oe] $T = 4.5$ K | $R$ | $M_S$ [emu/g] | $\Phi$ | $\mu_{eff}$ [$\mu_B$] |
|---|---|---|---|---|---|---|---|---|---|---|---|
| 0.185 | 0.060 | --- | 41 | 10 | 21.5 | 0.3 | 9 | 0.033 | 6.64±0.05 | 1.7±0.08 | 7.69 |
| 0.386 | 0.020 | --- | 17.9 | 2.54 | --- | 0 | 0 | --- | 1.85±0.03 | --- | 1.95 |
| 0.406 | 0.072 | --- | 19.7 | 10.5 | 10.5 | 0.5 | 14 | --- | 5.95±0.04 | --- | 8.08 |
| 0.405 | 0.052 | --- | 23.6 | 6.8 | 7.1 | 1.4 | 70 | --- | 4.81±0.04 | 1.55±0.05 | 5.23 |
| 0.545 | 0.030 | --- | 25.9 | 2.85 | --- | 0 | 0 | --- | 2.64±0.04 | --- | 2.19 |
| 0.593 | 0.062 | 7 | 17.2 | 6.57 | --- | 1.3 | 30 | --- | 4.8±0.05 | --- | 5.05 |
| 0.641 | 0.086 | 12 | 17.8 | 11.1 | --- | 3.9 | 160 | --- | 7.3±0.05 | --- | 8.54 |
| 0.724 | 0.040 | 4.3 | 30.3 | 3.2 | --- | 0.1 | 23 | --- | 3.11±0.03 | --- | 2.46 |
| 0.797 | 0.072 | 6 | 15.3 | 7.26 | --- | 2.7 | 55 | --- | 4.8±0.04 | --- | 5.58 |
| 0.841 | 0.077 | 9 | 20.2 | 10.8 | --- | 5.5 | 160 | --- | 6.7±0.05 | --- | 8.31 |

Measurements of the ZFC and FC curves were obtained at different magnetic field magnitudes in the range, 10 Oe $\leq H \leq$ 200 Oe. The $M_{ZFC}(T)$ and $M_{FC}(T)$ data was collected during the heating after the cooling from the temperatures above $T_C$ to 2 K without the magnetic field and in the applied field, respectively. In all $M_{ZFC-FC}(T)$ dependencies (Fig. 1(g,h,i)) the rapid increase of the magnetization value with the decrease in temperature is well visible. However, at lower temperatures, the aggregate doping concentration of Sn+Mn ions into α–GeTe seems to reveal different magnetic features which are also illustrated in Fig. 1(g,h,i). The ZFC curve for the crystal with $x \approx 0.4$, $y = 0.052$ shows broad maximum at $H = 15$ Oe, presented in Fig. 1(g). The estimated temperature of the peak's maximum of the ZFC curves was obtained as 8 K at $H = 15$ Oe for this crystal. Also, the largest splitting of the ZFC and FC curves is observed for the sample with $x \approx 0.4$, $y = 0.052$ among the samples presented in Fig 1(g,h,i). Such a broad maximum together with significant difference between $M_{ZFC}$ and $M_{FC}$ at low temperatures has been previously discussed to represent a spin or cluster–glass (CG) magnetic ordering [42,43]. The temperature at which the difference between ZFC and FC curves becomes non-zero, $\Delta M = M_{FC}(T) - M_{ZFC}(T) \neq 0$, named $T_{Irr}$, characterizes an irreversible magnetization process [44,45], and was estimated as 10 K measurement performed at 15 Oe for the in Fig. 1(g). The $\Delta M$ (at the lowest temperature) and $T_{Irr}$ values for the crystal with $x = 0.4$ and $y = 0.052$ decrease as the magnetic field increases. The idea of the $\Delta M$ parameter and its relation with the freezing process or frustration of magnetic moments, which might lead transition to a SG or CG ordering [46], will be presented in more detail in the discussion section.

In Fig. 1(h), we present the $M_{ZFC-FC}(T)$ results for the GSMT alloys with intermediate doped crystal incorporating $x \approx 0.6$, $y = 0.062$. Here, the low temperature part of the ZFC curves exhibit a flattened shape compared to Fig. 1(g) with a slow decrease towards low temperature. It might suggest that further Sn+Mn doping develops a magnetic ordering which is different from the crystals with $x = 0.4$. However, this trend in $M_{ZFC-FC}(T)$ relations is still similar to the properties reported in $La_{1-x}Sr_xCoO_3$ alloys [43]. For this sample, the split of the ZFC and FC curves, $\Delta M$, increases towards lower temperature similar as for the results from Fig. 1(g). Here, the two processes; drop in irreversibility temperature towards lower values and squeezing of divergence between $M_{FC}(T)$ and $M_{ZFC}(T)$ at high fields signify that frustration of magnetic moments may lead to SG state. It is important to note that the irreversibility temperature is less than that of the maxima, $T_{Irr} \lesssim T_{max}$, in case of canonical SG however; it is not true for the current samples which might be a signature of the magnetic clusters [47,48]. In Fig. 1(i), the $M_{ZFC-FC}(T)$ results are presented for heavily-doped crystals with the $x \approx 0.8$ and $y = 0.072$. As the sample is cooled down, the ZFC and FC curves exhibit identical magnetization up to $T = 2$ K with minor indication of irreversibility. Compared to the results in Fig. 1(g,h) for the other two samples, the ZFC and FC curves show flat plateaus in which the cusp like shape disappears at all applied fields. The magnetic interactions in Fig. 1(i) reveals a long range FM ordering as obvious from the saturated magnetization curves. Unlike the results in Fig. 1(g,h) which might present mixed magnetic interactions due the presence of glassy magnetic state, the sample in Fig. 1(i) shows domain effects which are typical feature of FM phase. So far, the tuning of magnetic interactions in Fig. 1 was presented as a function of Mn content. However; the glassy magnetic region suggests the involvement of Sn cannot be disregarded in addition to Mn ions. For example the sample with $y = 0.072$ in Fig. 1(a) manifests a frustrated magnetic behavior whereas $y = 0.072$ in Fig. 1(c) shows FM-like ordering. These results show that the GSMT samples do not alter magnetic state owing to Mn content only. The complexity of magnetic interactions for 5 – 7 % Mn content is attributed to clusters of different sizes that could also be seen comparing the broadness of $\chi_{AC}(T)$. Unlike the typical SG state, the present samples establish clusters of frozen FM-like ordering which were not observed in these alloys before [28-30]. The samples indicating Mn clusters will be analyzed in great detail in the following sections. In addition, the FM ordering in Fig. 1 is of long range without any signatures of spin freezing. Further, for $x \sim 0.2$, $y = 0.06$ and $x \sim 0.4$ and $y = 0.052$, the $\chi_{AC}(T)$ results are shown in Fig. S3(a-d). The symmetric and sharp maxima of $\chi_{AC}(T)$ and magnetic interactions in Fig. S3(a-d) will be analyzed in discussion part.

The magnetic features of the GSMT crystals were extended to the measurements of the static magnetization as function of applied field, $M(H)$. In Fig. 2, $M(H)$ results are shown for the whole series of GSMT crystals obtained at $T = 4.5$ K. In Fig. 2(a), $M(H)$ data for $x \sim 0.4$, $0.02 \leq y \leq 0.072$ are shown. For these samples, the hysteretic aspect and saturation magnetization, $M_S$, appear to depend on the Mn content where the maximum $M_S$ is achieved for $y = 0.072$. As pictured in the inset of Fig. 2(a), $y = 0.072$ manifests an eminently narrow $M(H)$ curve and the smallest magnetization remanence, $M_R = 0.5$ emu/g, in comparison with the results in Fig. 2(b,c) (excluding the sample with $x \sim 0.6$ and $y = 0.03$ exhibiting zero coercivity). Also, the crystals with the small Mn doping are saturated at $H \geq 40$ kOe, whereas those with the highest Mn content did not achieve saturation even at $H = 90$ kOe. Further increase in Sn+Mn content produces a noticeable hysteresis loop and $M_R$ as shown in the insets of Fig. 2(b,c). Also, $M_S$ as a function of Mn content follows a similar course as described for the sample with $x \sim 0.4$. In addition to the hysteric features as a function of doping contents, the temperature dependency of remanence, $M_R(T)$, and coercivity, $H_C(T)$, are shown for the samples with the highest content of Mn, Fig. 2(d). At $T = 4.5$ K, the crystals with $x \sim 0.6$, $y = 0.086$ and, $x \sim 0.8$, $y = 0.077$ achieved $M_R = 5.5$ emu/g and 3.9 emu/g, respectively, although the $H_C$ values of 160 Oe are same for both the crystals. Moreover, the crystals, $x \sim 0.6, 0.8$ maintain their hysteresis at elevated temperatures of $T = 15$ K and 10 K, sequentially.

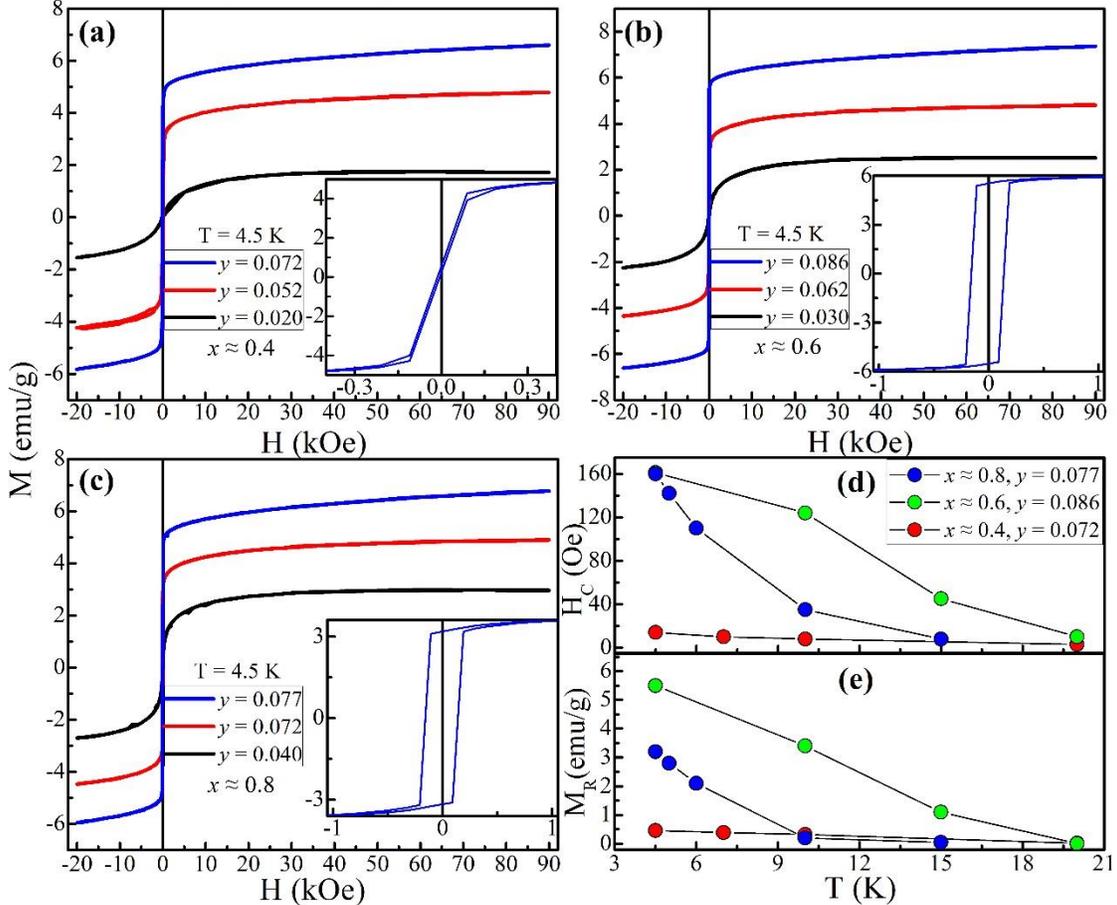

**FIG. 2**. $M(H)$ dependencies obtained at $T = 4.5$ K and $B \leq 90$ kOe for (a) $x \sim 0.4$, $0.02 \leq y \leq 0.072$ (b) $x \sim 0.6$, $0.03 \leq y \leq 0.086$ (c) $x \sim 0.8$, $0.04 \leq y \leq 0.077$. The $M(H)$ cuts are presented in the inset of (a,b,c) to depict the remanence and coercivity of crystals with highest Mn concentration in each figure. (d) Coercive field, $H_C(T)$ and (e) Temperature dependence of magnetization remanence, $M_R(T)$ for the same crystals.

For the sample $x \sim 0.6$ and $y = 0.062$, the $M(H)$ curve is shown in Fig. S4(a), thermal results of this sample were presented in Fig. 1(h). This is different from the square–like FM hysteresis in Fig. S4(b) and a characteristic S–shaped curve for SG state shown in Ref. 47. The $M(H)$ loop in Fig. S4(a) manifests an intermediate shape between a typical SG and FM hysteresis which might represent the presence of FM-like clusters in the sample as mentioned in Fig. 1(h). However; the $M_R(T)$ and $H_C(T)$ results show a sharp decrease above $T = 4.5$ K. Notably, the $M_R(T)$ and $H_C(T)$ values for $x \sim 0.4$, $y = 0.072$ are negligibly small even at the lowest measured temperature. Unlike the square-like FM hysteresis loops in Fig. 2(b,c), the extremely narrow S–shaped curve signifies a SG like magnetic state which could also be identified by the broad maximum in the $\chi_{AC}(T)$, see Fig. 1(a) [39,47].

## VI. DISCUSSION

In order to identify the nature of magnetic ordering and its variation as a function of doping, the $\chi_{AC}(T)$ and DC magnetization results are analyzed using several approaches. We begin the analysis by considering the crystals showing cusps in $\chi_{AC}(T)$ which is typically a signature of a transition to various magnetic states Fig. 1(a) [38,39]. For this reason, AC magnetic field with varying frequency is employed to understand the response of $\chi_{AC}(T)$. For SG like ordering, it has been widely investigated that the maximum of the cusp which represents a freezing temperature, $T_F$ in SG systems, moves up [28,49] on temperature axis as the AC magnetic field frequency increases. The Re($\chi_{AC}(T)$) dependencies collected at several $f$ values for the crystals with $x \sim 0.4$, $y = 0.052$ and $x \sim 0.2$, $y = 0.06$ are presented in Fig.S3 (a–d). The crystal with $x \sim 0.2$, $y = 0.06$ displays a prominent double maxima in $\chi_{AC}(T)$, Fig. S3(a,b). Here, the first maximum named $T_{F1}$ appears at $T \sim 10.6$ K of the Re($\chi_{AC}(T)$) data which is independent of the variation in frequency between $40 \le f$ (Hz) $\le 10000$. At $T \sim 21.5$ K, a broader and frequency dependent second maximum, $T_{F2}$ is recorded that moves to $T \sim 25$ K at $f = 10000$ Hz. The Im($\chi_{AC}(T)$) component of this sample also manifests double maxima at $T_{F1} \sim 7.5$ K and $T_{F2} \sim 15.1$ K that shows same frequency dependence as the Re($\chi_{AC}(T)$). Similar to double maxima, Eftimova et al., had observed shoulder–like shape in Im($\chi_{AC}(T)$) and its variation with applied frequency for $Pd_8Co_{50}Al_{42}$ SG alloy [47]. Also, Koyano et al., showed that a shoulder in both Re($\chi_{AC}(T)$) and Im($\chi_{AC}(T)$) parts occurred in $Fe_{1/4}TiS_2$ CG alloy which flattened with higher excitation frequency [42]. This occurrence of a shoulder in $\chi_{AC}(T)$ was attributed to the possible presence of small magnetic clusters in the material [42]. Similar to the previous works, the appearance of double maxima in the Re($\chi_{AC}(T)$) relation might also originate from the presence of magnetic clusters of varying sizes. The nature of magnetic state whether constituted of individual or collective behavior of spins will be further discussed in the following sections. Since the $T_{F1}$ peaks in both the Re($\chi_{AC}(T)$) and Im($\chi_{AC}(T)$) of Fig. S3(a,b) are frequency independent similar to FM phase, it probably represents magnetic clusters with FM like ordering at low temperature. The $T_F$ of the sample in Fig.S3 (c,d) however; did not show any frequency dependence up to $f = 10000$ Hz that could arise due to clustering of FM like ordering.

The cusps of Re($\chi_{AC}(T)$) for each $f$ value were first fitted with a polynomial function to estimate the accurate $T_F$ temperature. For $y = 0.06$, the $T_F$ value rises continuously in the frequency range $f = 44 - 10000$ Hz, the scaling parameter, $R$, for this crystal was assessed by using eq. 3 only for extreme frequencies.

$$R = \frac{\Delta T_f}{T_f (\log \Delta f)} \quad (3)$$

Here $\Delta T_f$ expresses the difference over which the peak temperature moves for a decade of frequencies, $f_1$ and $f_2$ and $\Delta f = f_2 - f_1$ defines the difference in two outermost applied frequencies. The value of scaling parameter, $R$, characterizes a particular type of magnetic ordering in the Re($\chi_{AC}(T)$) analysis. The $R$ value was calculated for extreme frequencies which resulted in $R = 0.033$ indicating the presence of a magnetic cluster state in the crystal. The values of $R$ for CG are typically ranked between 0.02 and 0.06 whereas it ranges from 0.005 to 0.01 for SG [50]. The nature of magnetic state and the presence of clusters in this sample will be further discussed in detail using phenomenological laws.

For $y = 0.052$ showing signs of glassy magnetic state as deduced from the $\chi_{AC}(T)$ shape, the evolution of the ZFC maximum visible in the results of the $M_{ZFC}(T)$ measurements as presented in Fig. 1(g). The maxima in ZFC curves together with the $T_{Irr}$ temperature move to lower temperatures as magnetic field is increased from 15 Oe to 200 Oe. Here, the estimated temperature values of maxima are marked by arrows of same colors.

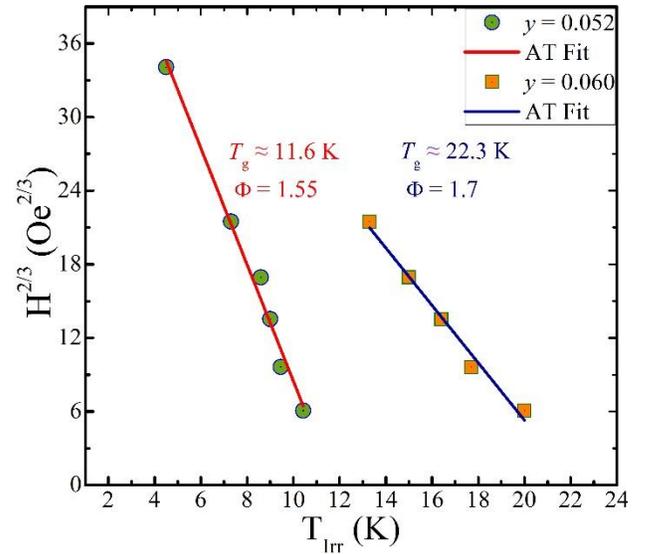

FIG. 3. The de Almeida-Thouless line fits (solid lines) plotted for $x \sim 0.2$, $y = 0.06$ and $x \sim 0.4$, $y = 0.052$ following the shift in $T_{Irr}$ values as a function of applied magnetic field. Based on AT-line, the lower temperature, $T$, and applied field, $H$, region illustrates a SG state which transforms to a PM state at high, $T$ and $H$, values.

In similar works on glassy magnetic states, the shift in $T_{Irr}$ to lower temperature with increasing $H$ is shown to follow a power law, the de Almeida-Thouless line [44,45] which follows the relation, $\delta T_{Irr} \propto H^{2/3}$. Here, $\delta T_{Irr}$ represents a shift in irreversibility temperature whereas $H$ denotes the applied magnetic field. The critical AT lines in Fig. 3 which conventionally separate a SG and PM state were obtained for the crystals after fitting the experimental results with eq. 4.

$$H = H_0 \left[1 - \left\{\frac{T_{Irr}(H)}{T_{Irr}(0)}\right\}\right]^{\Phi/2} \quad (4)$$

The best fits obtained in Fig. 3 show that the crystals follow the AT line in an H-T plane. Here, the field amplitude $H_0$, SG freezing temperature $T_{Irr}(0)$ or $T_g$ at $H = 0$, and crossover exponent $\Phi$ were used as free parameters. The values of $\Phi = 1.55 \pm 0.05$ and $1.7 \pm 0.08$ for $y = 0.052$ and $y = 0.06$ were obtained which show slight deviation from the standard values of $\Phi = 3/2$ for SG systems [44,45]. The obtained value of $\Phi = 1.55$ is close to $\Phi = 3/2$ for SG whereas $\Phi = 1.7$ for $y = 0.06$ is proximate to the CG system agreeing with the earlier work by Pallab et al., [51]. However; the small deviation for $y$ = 0.06 might also indicate that the glassy state either belongs to a non-mean-field universality class or arises due to strong anisotropy as discussed in earlier works [51,52]. The fitted AT lines shown here could be extrapolated towards the temperature scale which define the estimated SG/CG temperature, $T_g$ values at $H$ = 0. The $T_g$ values estimated during AT fitting were $T_g$ = 11.6 K and 22.3 K, for $y$ = 0.052, and 0.06, sequentially. Comparing these values with the freezing temperature estimated by Re($\chi_{AC}(T)$ as presented in Table II, the deviation seen for both crystals is assumed to arise due to possible uncertainty during the fitting.

Moreover, the $\Delta M$ term for $y$ = 0.052 persists to have non-zero values up to an applied field of 100 Oe. This reveals that the inception point of spin or cluster–glass state i.e. $\Delta M = M_{FC}(T) - M_{ZFC}(T) \neq 0$ is present up to 100 Oe. Upon further increasing the applied field to $H$ = 200 Oe in Fig. 1(g), the ZFC and FC curves follow the same pattern up to the lowest measured temperature where $\Delta M = M_{FC}(T) - M_{ZFC}(T) = 0$ which explains that the external magnetic field dominates the anisotropy of the sample. The decline of $T_{Irr}$ at high magnetic fields could also be explained by analyzing the AT lines in Fig. 3 which show the glassy magnetic ordering disappears in the high field region. Therefore, at high magnetic fields, the magnitude of $T_{Irr}$ could become negligible by extrapolating the fitted AT lines towards field-axis.

Based on the analysis of Fig. 1, the magnetic interactions alter from disordered locked-in state to a long range FM ordering as $y$ increases. It is important to see that the ZFC curves in Fig. 1(h) conclude at non-zero values below maxima unlike conventional glassy magnetic states. Such a disagreement might occur in the presence of weak FM like ordering within the glassy state. A similar non-zero ZFC curve was shown by Phan et al., for polycrystalline perovskite (Nd$_{0.65}$Y$_{0.35}$)$_{0.7}$Sr$_{0.3}$MnO$_3$ caused by the coexistence of FM and SG orderings [53]. For the glassy magnetic states in Fig. 1(g,h), the $M(B)$ hysteresis loops are shown in Fig. S5 and Fig. S4(a). The noticeable magnetic hysteresis and irreversible processes in $M(T)$ results are signatures of coexistent FM-like and glassy magnetic states in $x$ ~ 0.4, $y$ = 0.052 and $x$ ~ 0.6, $y$ = 0.062. For $x$ ~ 0.8, $y$ = 0.072, the sharp transition to an ordered phase and nearly indistinguishable/constant magnetization is generally attributed to the presence of FM phase [54].

In this section, the spin dynamics of the glassy magnetic state is investigated which provides a comprehensive understanding of the nature of magnetic interactions. And so, the variation in freezing temperature, $T_F$ as a function of frequency in Fig. S3 is first analyzed using Arrhenius law (AL) which is written as eq. 5 [55].

$$f = f_0 \exp\left(-\frac{E_a}{k_B T_F}\right) \quad (5)$$

The log($f$) vs 1/$T_F$ graph is shown fitted to AL for $x$ ~ 0.2, $y$ = 0.06 crystal where $f$ is the driving frequency in $\chi_{AC}$ results, and $\tau = 1/f$ denotes the dynamic fluctuation time. The activation energy or potential barrier, $E_a/k_B$ which defines a blockade between the two easy orientations [56] was used as a fitting parameter. The estimated value of $E_a/k_B$ is 385 K which is usually equal to about 1–2$T_F$ in classical SG, however; these are substantially higher for our samples using AL [55,57]. For other magnetically glassy systems, large values of barrier potential have been obtained in La$_{0.5}$Sr$_{0.5}$CoO$_3$ [58], CaBaFe$_{4-x}$Li$_x$O$_7$ [59], and U$_2$CuSi$_3$ [60]. Also, the single flip relaxation time which defines the shortest relaxation time existing in the system, $\tau_0$ [57] value was obtained for the same sample as 3.4 × 10$^{-9}$ s. The characteristic $\tau_0$ values for SG state normally occur in the range 10$^{-10}$ to 10$^{-13}$ s [51], which indicates the glassy magnetic state in present sample yields slightly larger spin relaxation time. Next, the $T_F$ dependence on frequency was analyzed using critical slowing down or power law (PL) described by eq. 6 to investigate the nature of the magnetic glassiness [55].

$$\tau = \tau_0 \left(\frac{T_F - T_g}{T_g}\right)^{-zv} \quad (6)$$

To define an appropriate fitting formula, the above equation was used in the form of eq. 7.

$$\ln(\tau) = \ln(\tau_0) - zv\ln\left(\frac{T_F - T_g}{T_g}\right) \quad (7)$$

Here, the ln($\tau_0$) term represents the intercept and $zv$ denotes the slope deduced from the linear fitting function. The SG temperature, $T_g$ used in this equation was estimated by extrapolation of the $T_F$ fit to $T_F(f = 0)$. The value of $\tau_0$ as shown in Fig. 4(b) was obtained as 7.9 × 10$^{-4}$ s which indicates the single spin flip time is significantly larger than the typical values for SG. The slow spin dynamics might indicate a possible contribution from the FM-like clusters in these samples [50,61,62]. The next important fitting parameter used in the PL fitting is dynamic critical exponent, $zv$, which illustrates the nature of spin dynamics and categorizes magnetic state. The deduced value by PL fitting is presented in Fig. 4(b) as $zv$ = 6.2 which lies in the range of pure SG i.e. $zv$ = 4 – 12 [47,63]. Since the PL fitting results into very large $\tau_0$ value compared to SG systems whereas $zv$ falls in the range defined for SG, therefore, it is assumed that the PL fitting yields unconvincing results to describe $T_F(f)$ dependency in the present sample. For this reason, the analysis is further extended in order to obtain conclusive parameters' values.

The nature of a magnetically glassy state was further analyzed by using Vogel-Fulcher (VF) law as shown in Fig. 4(c,d). The VF law used in Fig. 4 (c) and (d) is presented in the following form [55].

$$f = f_0 \exp\left(-\frac{E_a}{k_B(T_F - T_0)}\right) \quad (8)$$

In order to obtain a proper fitting form, eq. 8 was rearranged as below in the form of empirical VF law.

$$T_F = \frac{E_a/k_B}{\ln(f_0/f)} + T_0 \quad (9)$$

The data is plotted as $T_F$ vs 1/ln($f_0/f$) relation in Fig. 4(c) with fitted model described by eq. 9. The value of attempt frequency $f_0$ (=1/$\tau_0$) was fixed which was determined using AL fitting.

The potential barrier, $E_a/k_B$, and Vogel-Fulcher temperature, $T_0$, were put as fitting parameters. After the determination of $T_0$, the VF law was extended to estimate the values of single spin flip time, $\tau_0$ presented in Fig. 4(d). For this reason, the VF law was rearranged in the form of eq. 10 for suitable fitting.

$$\ln(f) = \ln(f_0) - \frac{E_a/k_B}{(T_F - T_0)} \quad (10)$$

The ln($\tau$) vs 1/($T_F - T_0$) data was fitted to eq. 10 with $f_0$ and $E_a/k_B$ as free parameters. By comparing the fitting parameters with other fitting laws in Fig. 4, the $\tau_0$ value was obtained as $5.6 \times 10^{-9}$ s which is comparatively smaller than reentrant SGs [64-66] however; similar to CG systems [47,51,67,68]. Similarly, the $E_a/k_B = 97 – 101$ K about 5 times the $T_F$ in Fig. 4(c,d) reveals a CG like behavior [58,59,67]. For canonical SG, the $E_a/k_B$ value remains $\sim 2T_F$ as previously reported for different systems [55,57].

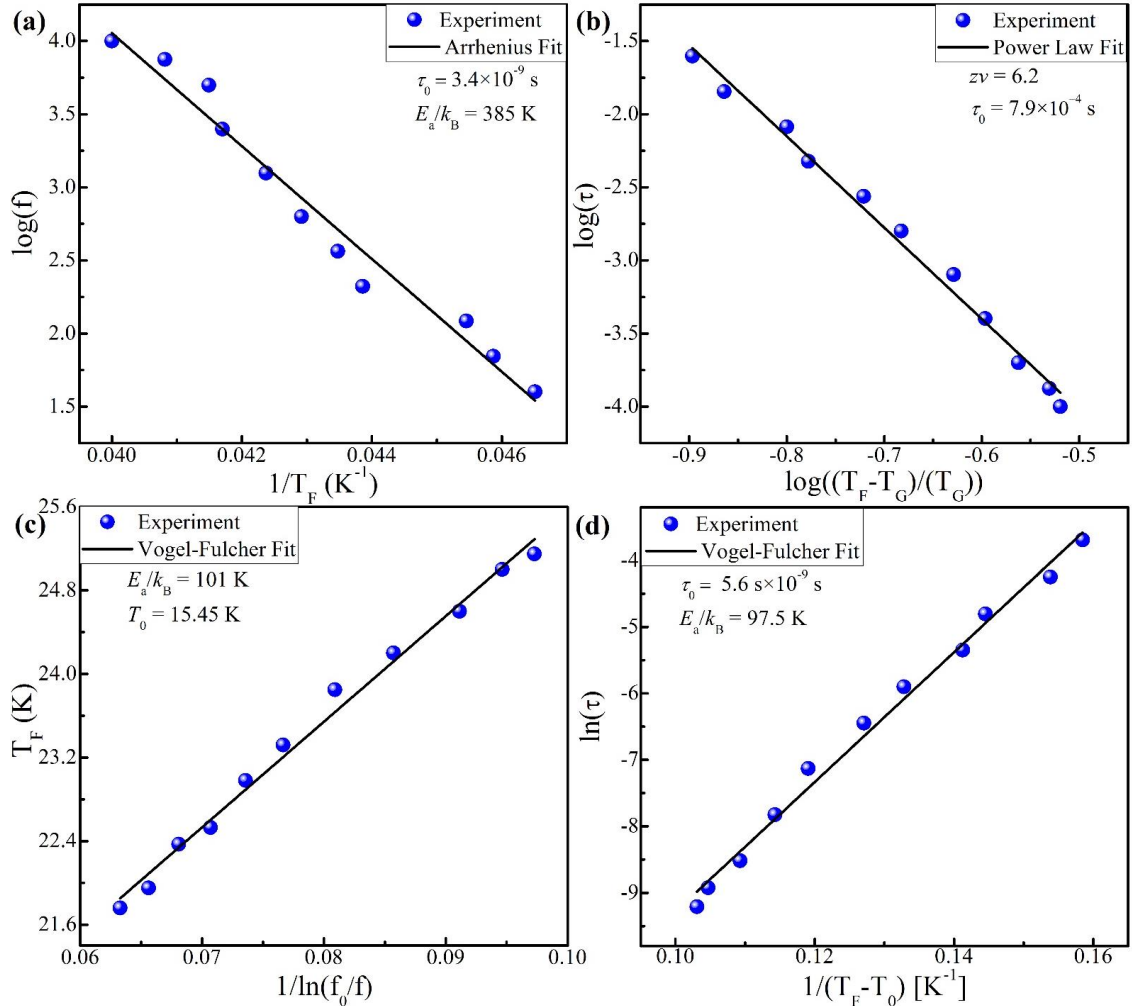

**FIG. 4**. The frequency dependence of the freezing temperature, $T_F$ analyzed with different models for magnetically glassy systems with $x \approx 0.2$, $y = 0.06$ (a) the variation in $T_F$ results was plotted as log($f$) vs $1/T_F$, the solid line represents fitting to the Arrhenius law. (b) The log($\tau$) vs log(($T_F$–$T_g$)/$T_g$)) data where $x$–axis represents reduced temperature; the result is fitted to the law of critical slowing down. (c) $T_F$ vs $1/\log(f_0/f)$, the solid lines characterize fits to empirical Vogel–Fulcher law. (d) The results are shown for log($\tau$) vs $1/(T_F$–$T_0)$ which are fitted to the Vogel–Fulcher law.

Furthermore, the VF temperature, $T_0$ obtained in Fig. 4(c) displays positive value which is a signature of interaction among the spins and development of magnetic clusters [68]. For $T_0$, the previous works suggest an estimated range between 0 K and $T_F$ [67], this implies that the empirical VF law provides convincing value of 15.45 K ($T_F$ = 21.5 K). The presence of clusters could also be identified by using a criterion introduced by Tholence which is described as $T^* = (T_F - T_0)/T_F$ [68,69]. The obtained value for $y = 0.06$ is $T^* = 0.28$ which is an order of magnitude higher than the values for SG reveals more evidence of magnetic clusters in the sample [67-69]. Besides the distinction between magnetic glassiness, $T_0$ values could be used to identify the nature of interaction among the magnetic moments. For $y = 0.06$, the determined value, $T_0 = 15.45$ K is close to its $T_F$ value which suggests that the sample has dominant RKKY interaction [68]. With the spin relaxation time, $\tau_0$ slightly larger than the canonical SG state implies that the magnetic moment originates from a cluster of spins rather than an atomic spin [70]. Also, the size of the clusters is assumed to grow as the temperature is reduced [62]. Moreover, the formation of clusters could be supported by the obtained values of VF temperature, $T_0$ which lies in the range of $T = 0$ K and $T_F$. Similarly, the activation energy, $E_a/k_B \sim 5T_F$ signifies the existence of collective behavior of spins or CG in the sample. Concluding the above discussion, the Re($\chi_{AC}$)($f$) results yielding the spin relaxation time, $\tau_0$, activation energy, $E_a/k_B$ and Tholence criterion favor the presence of magnetic clusters. Consequently, the glassy magnetic state in the sample is suggested to originate from magnetic clusters in coexistence with FM-like interactions.

The deduced parameters using eq. 5 – 10 were analyzed to draw comparison with previously reported works. In Fig. 5, the $\tau_0$ values for $y = 0.047$ and $0.06$ are marked as current work which represent a SG and CG state, respectively. Much slower spin dynamics were reported for reentrant spin-glasses in different compositions that holds the upper part in Fig. 5. The sample $y = 0.06$ with $\tau_0 \sim 10^{-9}$ s might represent magnetic moments of small cluster sizes as the obtained value is slightly larger than typical SG. Since the acceptable values of $\tau_0$ slightly vary in literature, therefore the boundaries separating different magnetic states might not seem as sharp as presented in Fig. 5. For example, the spin cluster-glass with $\tau_0 \sim 10^{-11}$ s in ref. 47 is illustrated as SG in other materials systems [51].

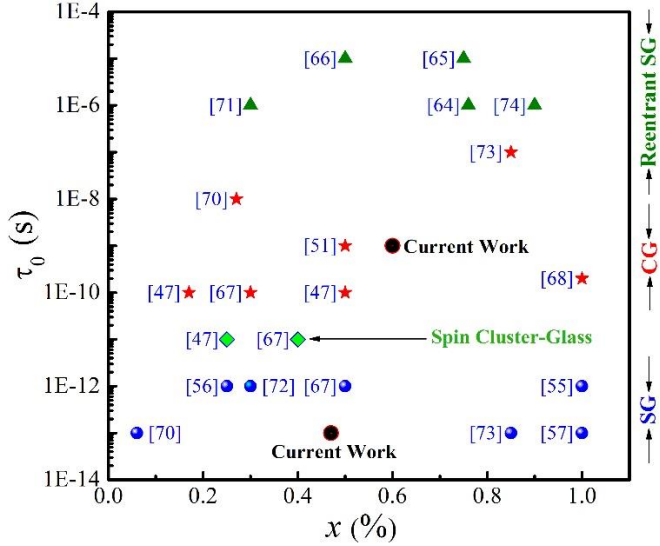

**FIG. 5**. Comparison of spin relaxation time, $\tau_0$ with the previous works [47,51,55–57,64–68,70–75]. The samples $x \sim 0.2$, $y = 0.047$ and $x \sim 0.2$, $y = 0.06$ represent a spin-glass and cluster-glass, respectively. Note that the horizontal axis denotes arbitrary scale for several references.

## VII. MAGNETIC PHASE DIAGRAM

The thorough magnetometric studies of GSMT crystals illustrate a variation in the nature of magnetic ordering for different Mn content. In this section, we present a magnetic phase diagram based on the above discussion. In the low doping region of $y \leq 0.04$, a negligible $\chi_{AC}(T)$ response was recorded exhibiting no signs of transition to an ordered phase up to the lowest measured temperature; see selected curves in Fig. 1(a–c), Fig. 2 and Fig. S6(a). Within this Mn range, the results manifest a spin disordered paramagnetic state in Fig. 6. The PM behavior of these crystals obviously signifies that there is no particular ordering of magnetic moments. As the Mn content is increased, the sample with $y = 0.047$ published earlier, demonstrated a SG magnetic ordering with a scaling parameter, $R = 0.017$ [30] shown as SG in Fig. 6. The hatched lines are approximate boundaries between different magnetic states. The ZFC–FC measurements were performed for this sample which shows a shift in $T_{Irr}$ as a function of magnetic field, Fig. S7. Also, the obtained values of $\tau_0 \sim 10^{-13}$ s and $zv = 4.9$ are in the range of canonical SG systems. For alloys with less than 5% Mn, as the $y$ content is increased from $0.2 - 0.4$ to $y = 0.47$, the magnetic moments freeze in a disordered state below the transition temperature. This sample establishes magnetic ordering which is different from the conventional long range ferromagnetic or antiferromagnetic (AFM) interactions. The disordered frozen state of SG system rather arises from a conflict between FM and AFM type interactions [76]. Moreover, the $\tau_0$ and $zv$ parameters also indicate that the $Mn^{2+}$ ions freeze as individual magnetic moments thus excluding the formation of clusters. Going further, the Mn content was increased from $y = 0.052$ to $0.06$ – shown as FM+CG in Fig. 6. As analyzed further using $T_F(f)$ results, the magnetic state behavior exhibits characteristics of Mn clustering. The magnetic moments show a locked-in disordered state however; different from the SG indicating Mn clusters. Since the CG state also manifests hysteresis loop, the clustering is assumed to constitute short range FM like ordering. As earlier reported by De et al., [77] and Chen et al., [78], the coexistence of FM phase with a glassy state originates by frozen FM spins. In the range, $\sim 0.52 \leq y \leq \sim 0.72$ the intracluster FM like interactions are presumed to play dominant part. Finally, samples with the highest Mn content in the range $0.072 \leq y \leq 0.086$ are presented. The Re($\chi_{AC}$) and $M(T)$ results showed a typical behavior of long range FM ordering in the samples. The FM ordering in current DMS is induced by holes via RKKY interaction. The spin-spin interactions between randomly distributed Mn ions in DMS are facilitated by delocalized or weakly localized holes.

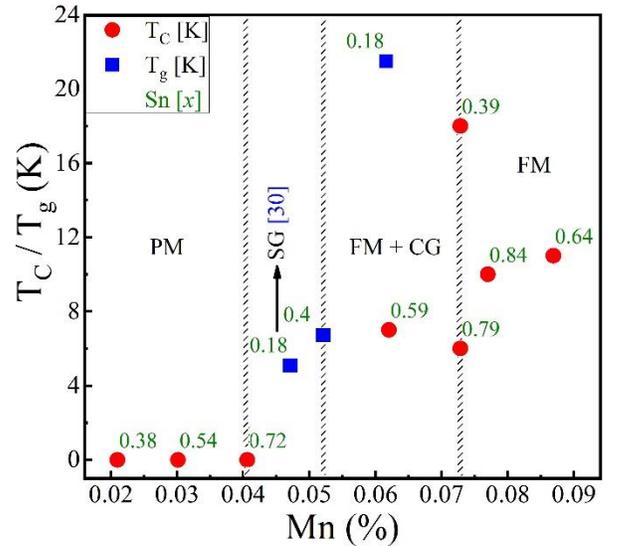

**FIG. 6**. Magnetic phase diagram of GSMT bulk crystals as a function of Mn content in the range, $0.021 \leq y \leq 0.086$. The spin–glass state for $y = 0.047$ denotes our previous work for the same quaternary alloy [30].

The overall magnetic behavior of the GSMT samples presents a distinct type of ordering which is different from our earlier work. Here the Mn clustering is attributed to excess of diamagnetic Sn ions along with Mn. As stated earlier, the presence of $Sn_{1-y}Mn_yTe$ lattice in Sn rich region plays vital role in initiating cluster formation. This also suggests that the magnetic phase diagram might look different with the same Mn content but in the low Sn region. Therefore, the overall impurity induced variation takes substantial contribution from Sn rich lattice, particularly, the CG region and beyond.

In order to interpret the response of the magnetic moments to the applied magnetic field, modified Brillouin function (BF) is generally used as a model to describe the PM state [79,80].

Since the crystals with low Mn contents reveal an inception to an ordered state at $T \approx 4.5$ K, therefore, BF for a FM was attempted to make approximate interpretation of $M(H)$ results. The $M(H)$ curves were fitted to eq. 11 which also yielded the saturation magnetization, $M_S$ values [79].

$$M = M_S + \chi_{dia} B \qquad (11)$$

Here, $M_S = g\mu_B N_0 S B_S(\zeta)$ where $B_S(\zeta)$ is modified Brillouin function,

$$B_S(\zeta) = \frac{2S+1}{2S} \coth\left(\frac{2S+1}{2S}\zeta\right) - \frac{1}{2S}\coth\left(\frac{\zeta}{2S}\right) \qquad (12), \text{ and}$$

$$\zeta = \frac{S g \mu_B (B + \lambda M)}{k_B T} \qquad (13)$$

In above equations, $g$ is known as the g-factor due to the magnetic $Mn^{2+}$ ion, $\mu_B$ represents Bohr magneton, $N_0$ is the number of cation sites in one gram of the composition, $S$ is spin of the $3d^5$ Mn ions which contribute $J_{Mn} = S_{Mn} = 5/2$, and $\chi_{dia} = -3\times10^{-7}$ emu/g is the diamagnetic contribution of the host lattice, GeTe. The best fits to $M(H)$ curves were obtained only for three samples as displayed in Fig. 7. The modified BF in this work was used with additional "$\lambda M$" term since the $M(H)$ curves were fitted in the FM region at $T \approx 4.5$ K. The added $\lambda M$ component in eq. 14 designates the molecular field term in the FM region [81]. The product $\lambda M$ denotes coefficient of the Weiss molecular field, $\lambda$ which describes the strength of the Weiss field and magnetization, $M$ due to molecular field, respectively. In ferromagnetic regime, the constant $\lambda > 0$ and molecular filed is substantially large due to sizeable coulomb energy associated with the exchange interaction [81]. For the BF interpretation in the PM region, the $\lambda M$ term in eq. 14 is excluded from the analysis. All of the crystals in Fig. 7 seem saturated at an applied field, $B > 40$ kOe as discussed earlier for Fig. 2.

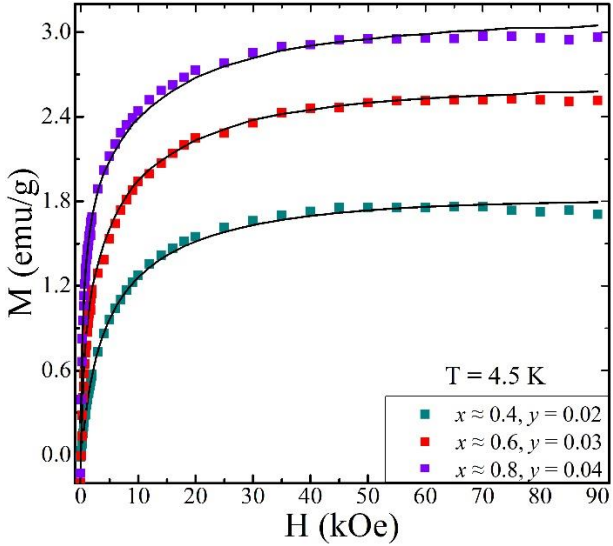

**FIG. 7**. $M(H)$ graphs of the selected crystals fitted with the modified Brillouin function for ferromagnets at $T \approx 4.5$ K. The samples $x$ = 0.38, 0.54, 0.72, $y$ = 0.02, 0.03, 0.04 are presented whereas solid lines show Brillouin fits.

Other than the interpretation of $M(H)$ curves with modified BF, the $M_S$ values were calculated where $M_S$ and $\lambda$ were put as free fitting parameters in the equation described above. The $M(H)$ results were initially tried with the BF without the $\lambda M$ term which did not yield the desired fits. However, as shown in Fig. 7, the fits obtained with the added FM term manifest reasonable interpretation of the low temperature magnetization. Here, $M_S$ obviously depends upon the doping content of Mn ions in the alloys indicating similar features as earlier $M(H)$ results presented in Fig. 2. Moreover, the second free parameter, $\lambda$ is crucial part of Weiss model that tells us about the nature of the alignment and interaction among the magnetic moments. Here, the positive values of Weiss constant, $\lambda$ = 1.27798, 1.0406 and 0.95 (g.Oe)/emu obtained for $y$ = 0.02, 0.03 and 0.04, respectively testify FM like ordering at $T$ = 4.5 K. It is therefore presumed that the proper fitting of the GSMT crystals is obtainable with Weiss model of BF and consequently reveals that the dominant exchange interactions between the $Mn^{2+}$ ions are positive.

Finally, we can discuss the obtained critical temperatures in terms of the strength of the RKKY interactions in the system. In order to properly take into account, the magnetic exchange in IV-VI DMSs such as GSMT, we used modified Sherrington – Southern (SS) model used by us previously for the same alloy. The detailed description of the SS model is described in Ref. 82 and results obtained for our previous GSMT samples with $x < 0.15$ are gathered in Ref. 29. The calculations were made with the exponential damping factor, $\lambda$ = 10 nm and the lattice parameters, $a$ = 5.98 Å and $\alpha$ = 88.8°.

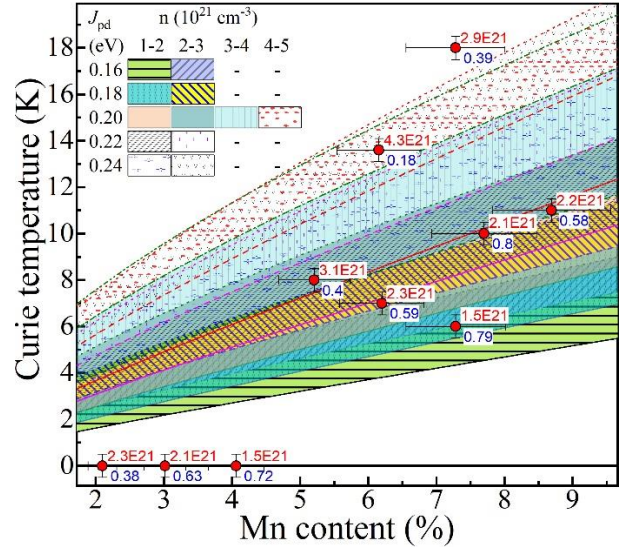

**FIG. 8**. Curie temperature, $T_C$, vs Mn content, $y$, obtained for the samples with different Sn contents, $x$ (marked as blue labels), and carrier concentrations, n (marked with red labels). Lines represent fits using modified Sherrington-Southern model for different magnetic exchange constant, $J_{pd}$, values and carrier concentration values.

As a result of our calculations, we obtained Curie temperature vs. Mn content dependence for several material parameters, such as Sn content, $x$, and Mn-hole exchange constant, $J_{pd}$. Experimental results together with the theoretical lines are presented in Fig. 8. The experimental results were reproduced well with rather low $J_{pd}$ values not exceeding 0.24 eV. We can also clearly see that the $J_{pd}$ value is showing a general decrease with the Sn content in the alloy from about 0.24 eV for the samples with $x \approx 0.2$ down to 0.16 eV for the samples with $x \approx 0.8$. The current values are lower than $J_{pd} \approx 0.45$ eV determined by us in Ref. 29 for GSMT with $x \approx 0.12$, in line with general trend illustrating decline in $J_{pd}$ vs Sn content. It is however striking that for the present case the Mn-hole exchange parameter does not change monotonically with Sn content, $x$, from the values reported for $Ge_{1-y}Mn_yTe$ equal to about 0.7 eV for bulk crystals [83] down to the value reported for $Sn_{1-y}Mn_yTe$ with $y > 0.1$ close to 0.1 eV [84]. The most important finding is related to the fact that the $J_{pd}$ vs $x$ dependence obtained in the present work are satisfying compared with the results presented in Ref. 29 and together these results show clearly that the $J_{pd}$ is not a linear function of the Sn content in IV-VI DMS alloys.

## VIII. CONCLUSIONS

The RT crystal structure of GSMT crystals manifests a doping dependent phase transition from polar to cubic phase. In the magnetometric part of the study, $Re(\chi_{AC})$ was extensively analyzed that yielded scaling parameter, $R = 0.33$ for $y = 0.06$ revealing magnetic clusters as compared to 0.017 of SG state [30]. In the doping range of $0.05 \leq y \leq 0.07$, the Mn ions accumulate into clusters of short range FM like ordering. The formation of Mn-clusters is attributed to the high proportion of Sn ions resulting from dissimilar lattice constants. Moreover, the ZFC–FC results were properly described with the AT fitting equation which showed deviation from the canonical SG state with $\Phi = 1.55, 1.7$ for $y = 0.052, 0.06$. These parameters, particularly the later one verify that clustering process of Mn ions is responsible for the glassy behavior. In order to validate this outcome, the barrier potential and spin relaxation time, $\tau_0 \sim 10^{-9}$ s which illustrates slower spin dynamics than the SGs proves clusters of magnetic moments. However; the $\tau_0$ value is rather close to SG range which elucidates that the clusters are of short range in this sample. Based on the $\chi_{AC}$ and ZFC–FC analysis, the interactions among Mn ions in the host lattice vary significantly depending on the doping content i.e. from PM to SG/CG to long range FM as presented in the phase diagram. Finally, modified SS model was used to understand the relation between Curie temperature and RKKY interaction in the samples. The obtained results of $J_{pd}$ exchange constant ranging from 0.16 to 0.24 eV showed a decrease with higher Sn content, in a good agreement with the previous results.

ACKNOWLEDGMENTS

The research was financed by the National Science Centre, Poland under the project number 2018/30/E/ST3/00309.
Corresponding author: akhaliq@ifpan.edu.pl

# Magnetic phase diagram of Ge$_{1-x-y}$(Sn$_x$Mn$_y$)Te multiferroic semiconductors: coexistence of ferromagnetic and cluster glass ordering

A. Khaliq,[1,a] S. Lewińska,[1] R. Minikaev,[1] M. Arciszewska,[1] A. Avdonin,[1] B. Brodowska,[1] V.E. Slynko,[2] A. Ślawska-Waniewska,[1] and L. Kilanski,[1]

[1]*Institute of Physics, Polish Academy of Sciences, Aleja Lotnikow 32/46, PL-02668 Warsaw, Poland*
[2]*Institute of Materials Science Problems, Ukrainian Academy of Sciences, Chernovtsy, Ukraine*

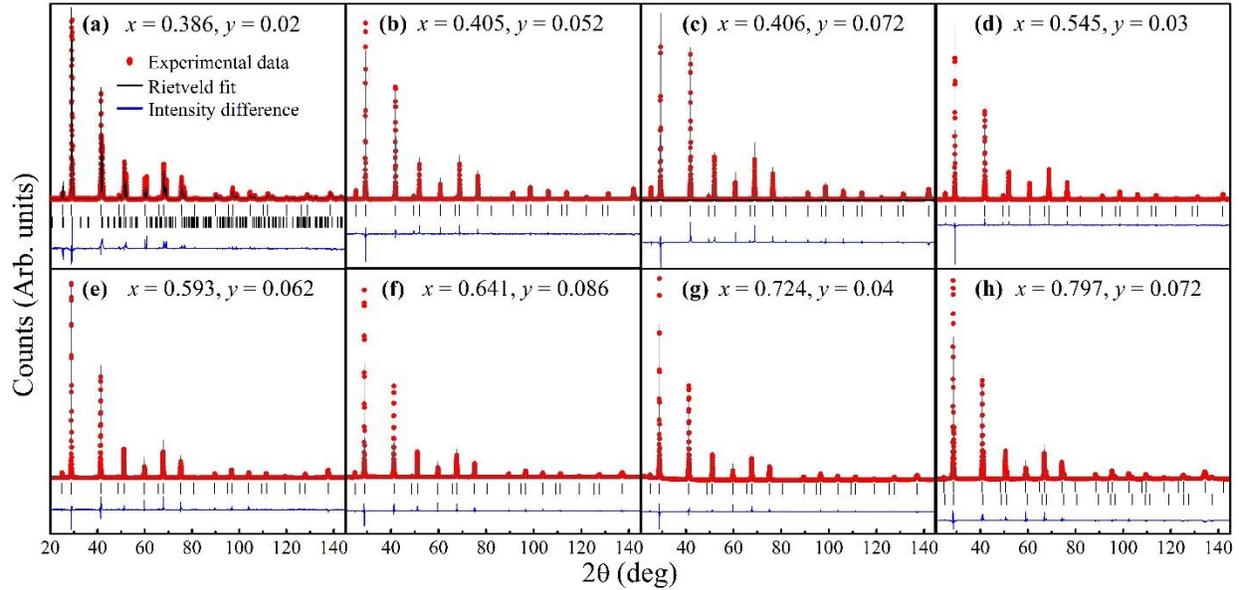

**Fig. S1** High resolution x−ray diffraction (HRXRD) results of Ge$_{1-x-y}$Sn$_x$Mn$_y$Te crystals in the doping range, $0.386 \leq x \leq 0.797$ and $0.02 \leq y \leq 0.086$. In (a), a mixed phase of rhombohedral and cubic symmetries was obtained indicating a boundary between the two phases, and (b–h) all the samples display a cubic phase for the high doping contents of Sn and Mn ions.

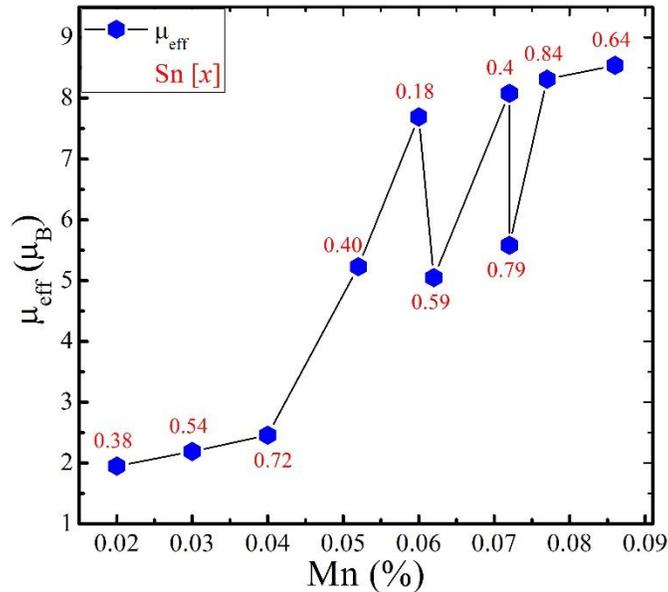

**Fig. S2** Effective magnetic moment of Ge$_{1-x-y}$Sn$_x$Mn$_y$Te bulk crystals with Mn doping content between $y = 0.02$ and $0.086$. The labels represent corresponding Sn content for each alloy. Except $y = 0.062$ and $0.072$, $\mu_{\text{eff}}$ is an increasing function of Mn content.

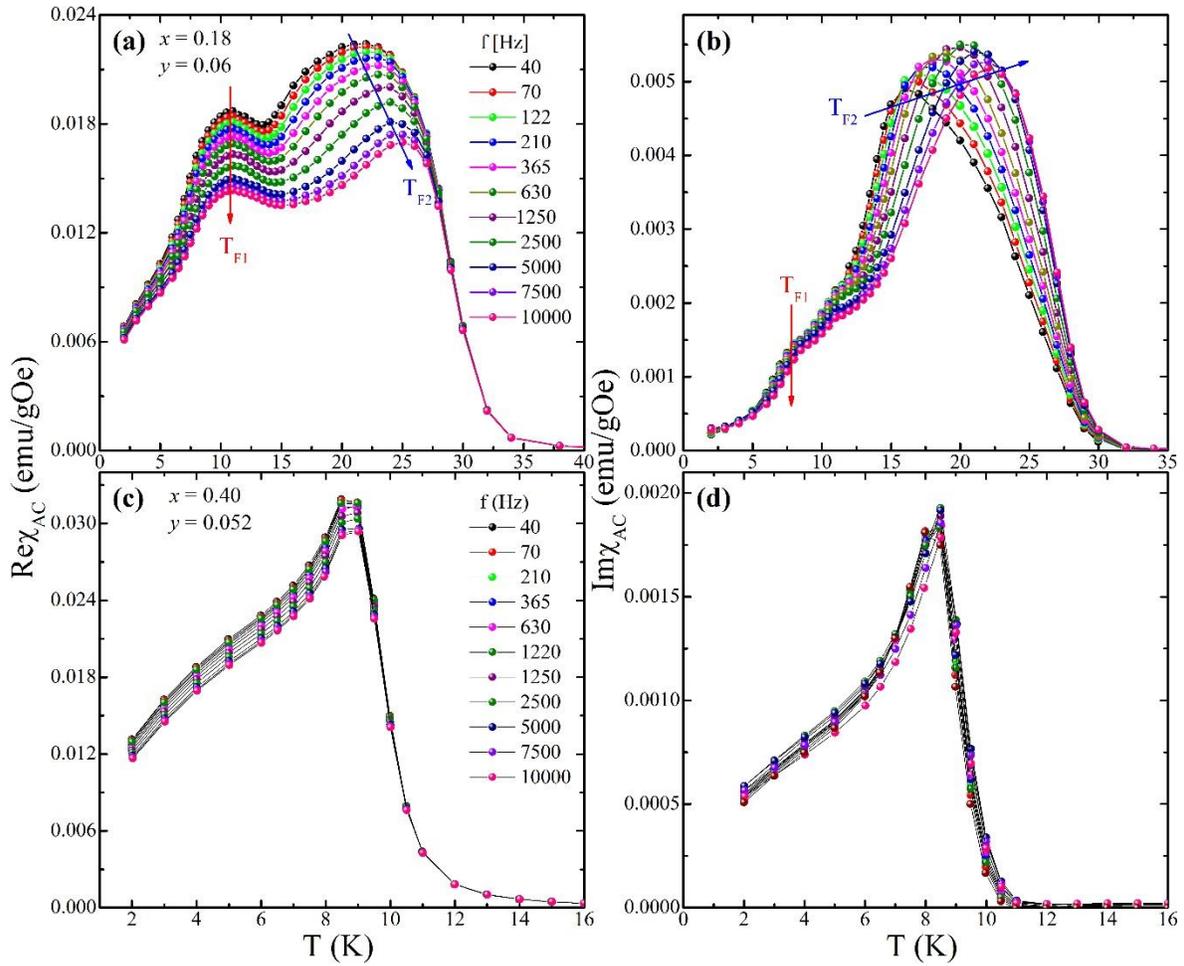

**Fig. S3 (a,b)** Re($\chi_{AC}$) and Im($\chi_{AC}$) components of the sample $x = 0.185$, $y = 0.06$ measured as function of frequency at $40 \leq f$ (Hz) $\leq 10000$. Both the Re($\chi_{AC}$) and Im($\chi_{AC}$) components display double maxima at $T_{F1} = 10.5$ K, 8 K and $T_{F2} = 21.5$ K, 15.5 K, respectively. (c,d) Re($\chi_{AC}$) and Im($\chi_{AC}$) components for $x = 0.40$, $y = 0.052$ measured under the same conditions.

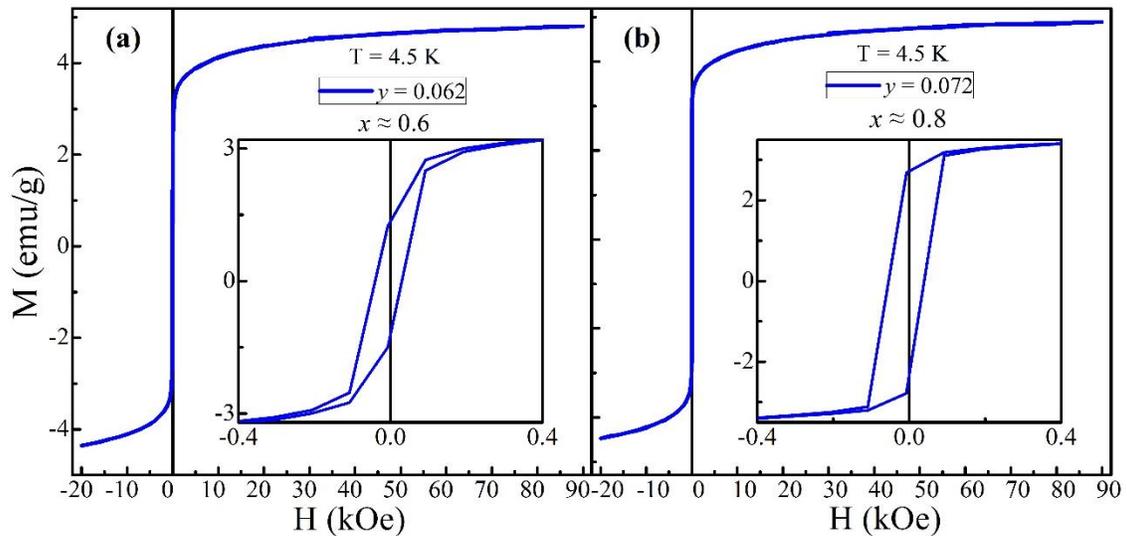

**Fig. S4 (a,b)** Magnetization hysteresis, $M(B)$ graph of $x \approx 0.6$, 0.8 and $y = 0.062$, 0.072, respectively. The insets show hysteresis cuts between, $-0.04 \leq B\,(T) \leq 0.04$ where (b) manifests a square–like hysteresis loop indicating a ferromagnetic ordering.

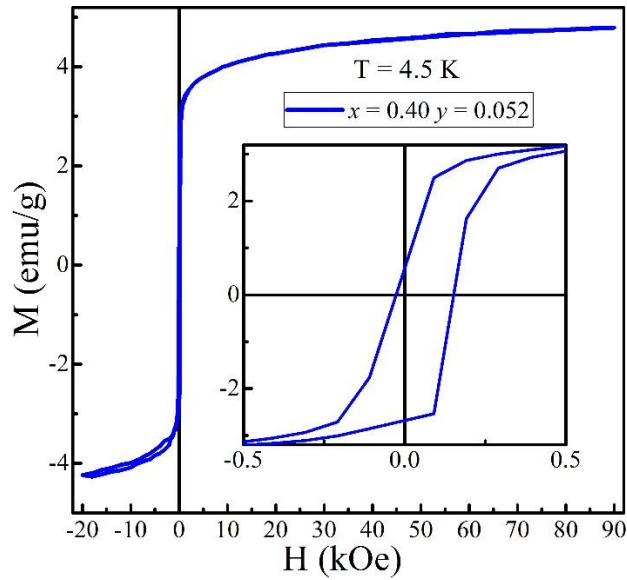

**Fig. S5** Magnetization hysteresis, *M(B)* graph for the sample with $x \approx 0.4$, $y = 0.052$. The inset shows noticeable spontaneous magnetization features of the sample indicating the presence of a ferromagnetic like ordering in coexistence with the glassy state.

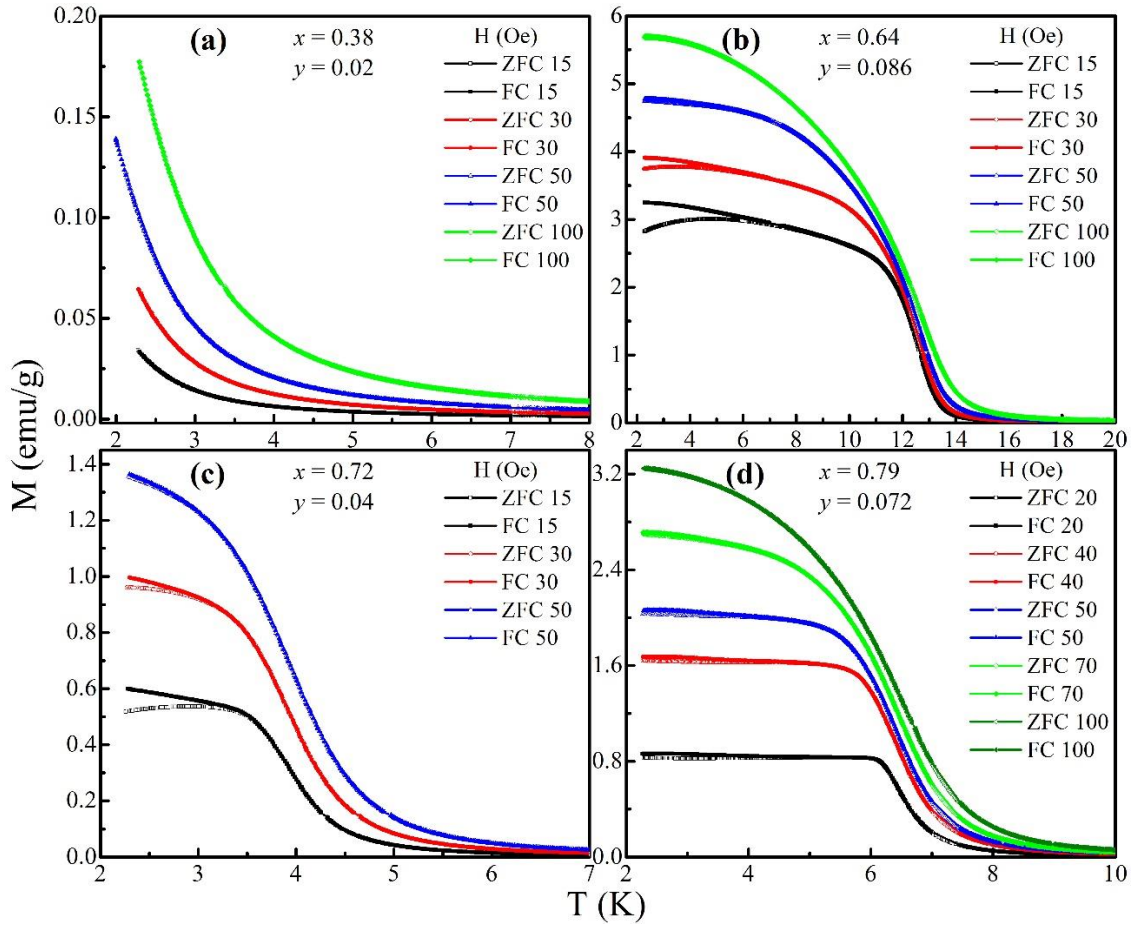

**Fig. S6** The zero−field−cooled and field−cooled (zfc−fc) dependence of $Ge_{1-x-y}Sn_xMn_yTe$ samples for (a) $x = 0.38$, $y = 0.02$ demonstrating characteristics a paramagnetic state, (b,c) samples with $x = 0.64$, $y = 0.086$ and $x = 0.72$, $y = 0.04$ displaying weak splitting in the zfc−fc curves which signifies the anisotropy of the system is higher than the external magnetic field, and (d) $x = 0.79$, $y = 0.072$ presents a negligible splitting in the fc-zfc graphs from $H = 20$ Oe to $H = 100$ Oe.

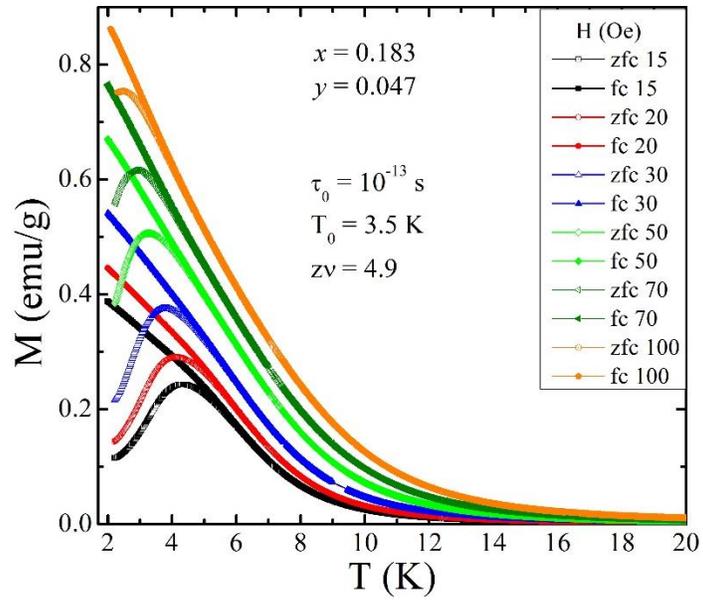

**Fig. S7** Magnetization, $M_{\text{fc-zfc}}(T)$ graphs of the sample $x \approx 0.2$, $y = 0.047$ illustrating a spin-glass magnetic state. The irreversibility temperature, $T_{\text{Irr}}$ shifts towards lower temperature as magnitude of the dc magnetic field is increased. The values of $\tau_0$, $T_0$ and $zv$ fall in the range of canonical spin-glasses.